\documentclass[12pt,titlepage]{article}
\usepackage[dvips]{graphicx}
\def\apm{{\it a priori} measure}
\newcommand{\me}{\mathrm{e}}
\newcommand{\dif}{\mathrm{d}}

\def\gst{g_s}
\def\abs#1{\left | #1 \right |}
\def\expval#1{\left\langle #1 \right\rangle}
\def\lambdar{\lambda_{r}}
\def\partialby#1{\frac{\partial\hfill}{\partial#1}}
\def\Dlambdar{\mathrm{D}\lambdar}

\def\dvol{\mathrm{dvol}}
\def\del{\nabla}
\def\Lambdazero{\Lambda_{0}}
\def\eff{\mathrm{eff}}

\def\betaeff{\beta_{\eff}}
\def\geff{g^{\eff}}
\def\aeff{a_{\eff}}
\def\ad{\gamma}
\def\Z{Z}
\def\rhor{\rho_{r}}
\def\Back{\mathrm{Back}}
\begin{document}
%
%
\begin{titlepage}
\hfill{\vbox{\hbox{hep-th/0212268}
 				\hrule width0pt height1.5\smallskipamount
 				\hbox{RUNHETC-2002-51}
 				\hrule width0pt height5\medskipamount}}
\vspace{12ex}
\begin{center}
{\Large Two talks on a
tentative theory of large distance physics}\\[10ex]
{\large Daniel Friedan}\\[4ex]
         Department of Physics and Astronomy\\
		 Rutgers, The State University of New Jersey\\
		 Piscataway, New Jersey, USA\\[1ex]
		 and\\[1ex]
		 Raunv\'{\i}sindastofnum H\'ask\'olans \'Islands\\
		 Reykjav\'{\i}k, \'{I}sland\\
		 The Natural Science Institute of the University of Iceland\\
		 Reykjavik, Iceland\\[2ex]
		 email: friedan@physics.rutgers.edu\\[12ex]
Talks given at the Carg\`ese 2002 Advanced Study Institute\\
\emph{Progress in string, field and particle theory}\\[2ex]
%
%
\end{center}
\end{titlepage}
\begin{abstract}
These talks present an overview of a tentative theory of large 
distance physics.  For each large distance $L$ (in dimensionless 
units), the theory gives two complementary descriptions of spacetime 
physics: quantum field theory at distances larger than $L$, string 
scattering amplitudes at distances smaller than $L$.  The mechanism 
of the theory is a certain 2d nonlinear model, the $\lambda$-model, 
whose target manifold is the manifold of general nonlinear models of 
the worldsurface, the background spacetimes for string scattering.  
So far, the theory has only been formulated and its basic working 
described, in general terms.  The theory's only claims to interest 
at present are matters of general principle.  It is a self-contained 
nonperturbative theory of large distance physics, operating entirely 
at large distance.  The $\lambda$-model constructs an actual QFT at 
large distance, a functional integral over spacetime fields.  It 
constructs an effective background spacetime for string scattering 
at relatively small distances.  It is background independent, 
dynamically.  Nothing is adjustable in its formulation.  It is a 
mechanical theory, not an S-matrix theory.  String scattering takes 
place at small distances within a mechanical large distance 
environment.  The $\lambda$-model constructs QFT in a way that 
offers possibilities of novel physical phenomena at large distances.  
The task now is to perform concrete calculations in the 
$\lambda$-model, to find out if it produces a physically useful QFT.
\end{abstract}
\newpage
These talks present an overview of a tentative theory of large 
distance physics \cite{tentative}.  This written version of the 
talks differs from the actual spoken version in its arrangement and 
in some of the content.  The transparencies that were used in the 
actual talks are available at \cite{transparencies}.  References can 
be found in \cite{tentative}.  No references will be given here.

\subsection*{The need to produce QFT}

My goal is a theoretical machinery capable of producing a definite, 
specific spacetime quantum field theory.

Everything, or almost everything, that is presently known about the 
laws of physics is summarized in the Standard Model of particle 
physics combined with General Relativity.  General Relativity is a 
classical field theory, but it can just as well be regarded as a 
cutoff quantum field theory which we observe only in its large 
distance, classical regime.  From this point of view, almost 
everything that is presently known about the laws of physics is 
summarized in one particular quantum field theory.  An enormous body 
of experimental data is summarized in this amazingly successful QFT.

Since the establishment of the Standard Model, theoretical 
speculation has lacked the guidance and correction of a rapid flow 
of new experimental results.  Speculative exploration of large 
theoretical search spaces with many free parameters could drag on 
forever without such guidance and correction.  We urgently need a 
theory that is definite and specific.  We need a theory that gives a 
definite and specific explanation of the Standard Model combined 
with General Relativity, so its reliability will be supported 
securely by the body of existing experimental knowledge.

To have a chance of explaining the Standard Model and General 
Relativity, a theory must first be capable of producing a {QFT}.  It 
is not enough merely to identify a QFT, say by matching to an 
S-matrix.  An S-matrix is not a mechanics.  An actual QFT must be 
constructed.  The theoretical machinery must produce a quantum 
mechanical space of states and a hamiltonian operator or, 
equivalently, a functional integral over spacetime fields, cut off 
perhaps at an unnoticeably small spacetime distance.

My strategy has been to look first for a well-defined theoretical 
machinery that is capable of producing a specific spacetime QFT, 
without assuming in advance that QFT operates.  The spacetime fields 
of the QFT should include the spacetime metric, some gauge fields, 
and some chiral fermion fields.  The dynamics should be generally 
covariant and gauge invariant.  But otherwise I have postponed 
worrying about the details of the QFT that is produced.  Until now, 
I have only been concerned with the problems of formulating a 
theory.  The task now is to figure out what the theory does, and 
eventually to check whether it does in fact produce the Standard 
Model plus General Relativity, in all known details.  If it fails to 
do so, then of course the theory is wrong.

\subsection*{The general nonlinear model}

The theory is based on the renormalization of the 2d general 
nonlinear model, which was the subject of my first talk at 
Carg\`ese, as a student in the summer of 1979.  The general 
nonlinear model is a functional integral over maps $x(z,\bar z)$ 
from the plane to a given compact, riemannian target manifold.  The 
action is
\begin{equation}
A(x)= \int \dif^{2}z \, \frac1{2\pi} \,
h_{\mu\nu}(x) \, \partial x^{\mu} \, \bar \partial x^{\nu}
\end{equation}
where $h_{\mu\nu}(x)$ is a positive definite, riemannian metric on 
the target manifold.  In local coordinates on the target manifold, 
the map $x(z,\bar z)$ is described by coordinate maps 
$x^{\mu}(z,\bar z)$.  Expanding the metric coupling in Taylor 
series, the action takes the perturbative form
\begin{equation}
A(x)= \int \dif^{2}z \, \frac1{2\pi} 
\, \left ( h_{\mu\nu} \, \partial x^{\mu} \, \bar \partial x^{\nu} + 
h_{\mu\nu,\sigma}x^{\sigma}
\, \partial x^{\mu} \, \bar \partial x^{\nu}
 +\cdots 
\right )
\end{equation}
which describes interacting massless scalar fields $x^{\mu}(z,\bar 
z)$.  The matrix $h_{\mu\nu}$ governs the gaussian fluctuations 
around the origin of coordinates $x^{\mu}=0$.  The propagator is
\begin{equation}
\expval{\,x^{\mu}(z_{1},\bar z_{1}) \, x^{\nu}(z_{2},\bar z_{2})\,}
= - h^{\mu\nu}
\, \ln (\mu^{2}\abs{z_{1}-z_{2}}^{2})
\end{equation}
where $\mu^{-1}$ is a reference 2d distance.  The higher Taylor 
series coefficients of the metric are the perturbative couplings of 
the model.  Free massless scalar fields are dimensionless in 2d, so 
all the couplings have naive scaling dimension 0.  The general 
nonlinear model is a renormalizable 2d field theory with infinitely 
many naively renormalizable coupling constants.

I called it the \emph{general} nonlinear model because the metric 
$h_{\mu\nu}(x)$ was not assumed to have any special symmetries and 
the renormalization of the model was generally covariant in the 
target manifold, and because the model contained all possible 
naively renormalizable couplings of the dimensionless scalar fields 
$x^{\sigma}(z,\bar z)$.

Renormalize the model at 2d distance $\mu^{-1}$.  The renormalized 
metric coupling follows the renormalization group equation
\begin{equation}
\mu\partialby\mu \, h_{\mu\nu}(x) = 
\beta_{\mu\nu}(x) = 2 R_{\mu\nu}(x) + \mathrm{O}(R^{2})
\end{equation}
where $R_{\mu\nu}(x)$ is the Ricci tensor of the metric.  The model 
is scale invariant when $h_{\mu\nu}(x)$ satisfies the fixed point 
equation $\beta=0$,
\begin{equation}
0 = 2 R_{\mu\nu} + 
\mathrm{O}(R^{2})
\end{equation}
The appearance of Einstein's equation (without matter) in this 
formal 2d setting, apparently unrelated to spacetime physics, seemed 
potentially a clue to an explanation of the physical Einstein 
equation.  It became possible to imagine that spacetime physics 
might be explained by some theoretical structure based on the 
renormalization of the general nonlinear model.  The target manifold 
would be spacetime.  But the equation $\beta=0$ was only the fixed 
point equation of the renormalization group flow in a 2d model.  It 
had the form of the spacetime field equation, but it was not the 
equation of motion of a mechanical system.

\subsection*{The coupling constants $\lambda^{i}$}

The renormalization group flow of the general nonlinear model 
exposed a direct relationship between the short distance properties 
of the 2d model and the large distance geometry of its target 
manifold.  Take a particular scale invariant model to serve as 
reference point.  The metric $h^{\mathrm{ref}}_{\mu\nu}(x)$ of the 
reference model solves the fixed point equation $\beta=0$.  A nearby 
model has metric $ h_{\mu\nu}(x) = h^{\mathrm{ref}}_{\mu\nu}(x) + 
\delta h_{\mu\nu}(x) $.  The scaling behavior of the perturbation 
$\delta h_{\mu\nu}(x)$ at short 2d distances is determined by the 
linearization of $\beta$,
a covariant second order differential operator
\begin{equation}
\beta_{\mu\nu} = (- \del^{\sigma} \del_{\sigma} + \cdots)
\, \delta h_{\mu\nu}(x) +\mathrm{O}(\delta h^{2})
\:.
\end{equation}
The perturbation can be expanded in eigen-modes
\begin{equation}
\delta h_{\mu\nu}(x) = \lambda^{i} \,\delta_{i}  h_{\mu\nu}(x)
\end{equation}
\begin{equation}
(- \del^{\sigma} \del_{\sigma} + \cdots )\; \delta_{i} h_{\mu\nu}(x)
= \ad(i) \;  \delta_{i} h_{\mu\nu}(x)
\end{equation}
which form a discrete set because the target manifold was taken 
compact and riemannian.  The summation convention is used for 
indices $i,j$ which range over the discrete set of wave modes.  Each 
eigen-mode corresponds to a local 2d scaling field
\begin{equation}
\phi_i(z,\bar z) =
 \mu^{-2} \, \delta_{i} h_{\mu\nu}(x) \,
\partial x^{\mu} \, \bar \partial x^{\nu}
\end{equation}
whose scaling dimension is $2+\ad(i)$.  The eigenvalue $\ad(i)$ is 
the anomalous dimension of $\phi_i(z,\bar z)$.  The perturbation of 
the action is
\begin{equation}
\delta A = \int \dif^{2}z \, \mu^{2} \frac1{2\pi} \,
\lambda^{i} \phi_i(z,\bar z)
\:.
\end{equation}
The perturbed model is made by inserting
\begin{equation}
\me^{-\delta A} = 
\me^{-\int \dif^{2}z \, \mu^{2} \frac1{2\pi} \,
\lambda^{i} \phi_i(z,\bar z)}
\end{equation}
into the reference model.  The wave modes $\lambda^{i}$ of the 
metric become the coupling constants that parametrize the 2d model.  
The renormalization group equation, written in terms of the 
$\lambda^{i}$, is
\begin{equation}
\mu\partialby\mu \: \lambda^{i} =
\beta^{i}(\lambda) = \ad(i) \, \lambda^{i}  + \cdots
\end{equation}
so the coupling constant $\lambda^{i}$ has dimension $-\ad(i)$.  The 
spectrum of dimensions $-\ad(i)$ is discrete and bounded above, 
because the target manifold is compact and riemannian.

Distances in the target manifold, which are to be spacetime 
distances, are pure numbers.  Implicitly, there is a basic unit of 
spacetime distance.  A distance is large when it is a large number 
in dimensionless units.  The target manifold is \emph{macroscopic} 
when it is very large in some of its dimensions, containing large 
regions where spacetime curvature and topology are negligible in the 
macroscopic dimensions.  Locally in a macroscopic target manifold, 
the wave modes $\lambda^{i}$ are approximated by plane waves with 
spacetime wave vectors $p(i)$.  The anomalous dimensions, being the 
eigenvalues of a second order differential operator, are essentially
\begin{equation}
\ad(i) = p(i)^{2}
\:.
\end{equation}
The eigenvalues $\ad(i)$ become tightly packed near 0, approximating 
a continuus spectrum indexed by the wave vectors $p(i)$.

The wave modes of the spacetime metric are the coupling constants 
$\lambda^{i}$ of the general nonlinear model.  The anomalous 
dimensions $\ad(i)$ of the 2d scaling fields are geometrical 
quantities in spacetime, essentially the squares of the spacetime 
wave vectors $p(i)$.  A dictionary between spacetime physics and the 
general nonlinear model begins to take form.

\subsection*{At short 2d distances $\Lambda^{-1}$}

From now on, the reference 2d distance $\mu^{-1}$ at which the 
general nonlinear model is renormalized will be held fixed.  
Everything will happen at much shorter 2d distances $\Lambda^{-1}$, 
where the renormalized model is parametrized by running coupling 
constants $\lambda_{r}^{i}$,
\begin{equation}
\me^{-\int \dif^{2}z \, \mu^{2} \frac1{2\pi} \,
\lambda^{i} \phi_i(z,\bar z)}
=
\me^{-\int \dif^{2}z \, \Lambda^{2} \frac1{2\pi} \,
\lambdar^{i} \, \phi^{\Lambda}_i(z,\bar z)}
\:.
\end{equation}
The $\phi^{\Lambda}_i(z,\bar z)$ are the scaling fields normalized 
at the short 2d distance.  The $\lambda_{r}^{i}$ run with 
$\Lambda^{-1}$ according to the renormalization group equation
\begin{equation}
\Lambda\partialby\Lambda \: \lambda_{r}^{i} =
\beta^{i}(\lambda_{r}) = \ad(i) \, \lambda_{r}^{i}  + \cdots
\end{equation}
whose integrated flow
\begin{equation}
\lambdar^{i} = (\Lambda/\mu)^{\ad(i)} \, \lambda^{i}
+\cdots
\end{equation}
when expressed in terms of the large number
\begin{equation}
L^{2} = \ln (\Lambda/\mu)
\end{equation}
shows the exponential suppression
\begin{equation}
\lambda^{i} = \me^{-L^{2}\ad(i)} \, \lambdar^{i}
            = \me^{-L^{2}p(i)^{2}} \, \lambdar^{i}
\:.
\end{equation}
When $\ad(i)$ is positive and large on the scale set by $L^{-2}$, 
the short distance coupling constant $\lambdar^{i}$ is 
\emph{irrelevant}.  It has no significant effects in the 
renormalized model.  The corresponding $\lambda^{i}$ has dimension 
$-\ad(i)$ which is significantly negative on the scale set by 
$L^{-2}$.  Such coupling constants $\lambda^{i}$ are 
\emph{non-renormalizable}.  The non-renormalizable coupling 
constants play no role in the renormalized model.  The 
renormalization puts them to zero.  Only the \emph{renormalizable} 
coupling constants matter in the renormalized model, whose 
dimensions $-\ad(i)$ are either non-negative or only slightly 
negative, the latter meaning close to zero on the scale set by 
$L^{-2}$.  All significant perturbations of the renormalized model 
at the short 2d distance $\Lambda^{-1}$ are parametrized by the 
renormalizable $\lambda^{i}$, or, equivalently, by the 
$\lambdar^{i}$ modulo the irrelevant $\lambdar^{i}$.

To be specific, say that $\lambdar^{i}$ is irrelevant if $ 
\me^{-L^{2}\ad(i)} < \me^{-400} $, the number $400$ being more or 
less arbitrary.  This is equivalent to $ \ad(i) > 400/L^{2} $ or $ 
p(i) > 20/L $.  Then the renormalizable coupling constants 
$\lambda^{i}$ are those with $\ad(i) \le 400/L^{2}$, which is $p(i) 
\le 20/L$.  The renormalizable coupling constants at the short 2d 
distance $\Lambda^{-1}$ are the spacetime wave modes at spacetime 
distances larger than $L$.  The number $L$ acts as an ultraviolet 
cutoff distance in spacetime.  It is a strong UV cutoff in the sense 
that the spacetime wave modes at distances smaller than $L$ are 
decoupled from the renormalized 2d model.  There is no significant 
dependence on the irrelevant $\lambdar^{i}$.  The decoupling is 
accomplished in the renormalization of the model, when the 
non-renormalizable $\lambda^{i}$ are forced to 0.

The large distance structure of spacetime is now translated into the 
short distance structure of the 2d model.  Specifically, the general 
nonlinear model at 2d distances shorter than $\Lambda^{-1}$ 
represents the mechanical structure of spacetime at distances larger 
than $L$.

The distinction between renormalizable and non-renormalizable 
coupling constants is familiar.  In a typical model, all the 
negative dimension coupling constants are non-renormalizable.  Once 
the ratio $\Lambda/\mu$ is sufficiently huge, it becomes effectively 
infinite.  Only the zero and positive dimension coupling constants 
are renormalizable.  All effects of irrelevant coupling 
constants at short distance are absorbed into shifts of the 
renormalizable coupling constants.  The renormalized model is 
parametrized entirely by the fixed set of renormalizable coupling 
constants.  The general nonlinear model is unusual in that its 
spectrum of scaling dimensions $-\ad(i)$ can crowd arbitrarily close 
to zero when its target manifold becomes arbitrarily large.  The set 
of renormalizable coupling constants depends on the ratio 
$\Lambda/\mu$, no matter how huge that ratio becomes.

\subsection*{The manifold of general nonlinear models}

The renormalizable coupling constants $\lambda^{i}$ serve as local 
coordinates parametrizing the renormalized models in the vicinity of 
the reference model.  The collection of all such local coordinate 
systems defines the manifold of renormalized general nonlinear 
models, $M(L)$.  It depends on $L$ because the set of renormalizable 
$\lambda^{i}$ depends on $L$.  Locally, $M(L)$ is the manifold of 
spacetime fields at spacetime distances larger than $L$.  Locally, 
$M(L)$ is finite dimensional, because the spectrum of $\ad(i)$ is 
discrete and bounded below, which is because the target manifold, 
spacetime, is compact and riemannian.  But the dimension of $M(L)$ 
grows without bound when the spacetime becomes large.

In the idealized limit $\Lambda^{-1}=0$, $L=\infty$, the only 
coupling constants that survive are the relevant coupling constants 
(positive dimension) and the marginal coupling constants (zero 
dimension).  The marginal coupling constants are the $\lambda^{i}$ 
that preserve scale invariance, at least infinitesimally.  They 
parametrize the manifold $M(\infty)$ of scale invariant models.  The 
marginal $\lambda^{i}$ are the zero modes, $\ad(i)=0$, in spacetime.

\subsection*{The general nonlinear model in string theory}

The general nonlinear model found use in string theory, to construct 
the string worldsurface in a curved background spacetime.  The 
general nonlinear model provides the local structure of the 
worldsurface.  The target manifold is the background spacetime in 
which strings scatter.  String theory uses somewhat elaborate 
versions of the general nonlinear model, incorporating 2d 
supersymmetry and 2d chiral asymmetry.  The couplings of the model 
are the spacetime metric and also some additional spacetime fields, 
which include spacetime gauge fields, fermionic spacetime spinor 
fields and more.  The $\lambda^{i}$ are the wave modes of this 
entire collection of spacetime fields, all the coupling constants of 
the general nonlinear model of the worldsurface.

The perturbative algorithm for calculating the string S-matrix needs 
the worldsurface to be scale invariant.  The equation $\beta=0$ 
again looks like a spacetime field equation, but it is still not the 
equation of motion of a mechanical system.  It is only a consistency 
condition on the background spacetime.  At low momenta and energies, 
the perturbative string S-matrix can be identified with the 
perturbative S-matrix of a spacetime {QFT}.  The classical field 
equation of the QFT can be identified with the consistency condition 
$\beta = 0$.  These correspondences strongly suggested that there 
should be a mechanical means of producing an actual QFT with $\beta 
= 0$ as its equation of motion, and that the means of production of 
the QFT should be related to the construction of the background 
spacetime for string scattering.  To find such a means of producing 
an actual {QFT} has been an urgent problem for many years.

Take the heterotic string theory, to be specific.  $M(L)$ is now the 
manifold of general nonlinear models of the heterotic worldsurface.  
The $\lambda^{i}$ are the coupling constants of the heterotic 
worldsurface (after GSO projection), which are the spacetime wave 
modes of the spacetime metric, gauge fields, scalar fields, 
antisymmetric tensor field, and chiral fermion fields.  Some of the 
$\lambda^{i}$ are fermionic, so $M(L)$ is now a graded manifold.  
The absence of tachyons means that $ \ad(i) \ge 0 $ for all 
$\lambda^{i}$.  There are no relevant coupling constants.  The 
renormalization group flows inward to the manifold $M(\infty)$ of 
exactly scale invariant models, the exact solutions of $\beta=0$.  
$M(\infty)$ is essentially the manifold of 10d compact Calabi-Yau 
target spaces (with some additional geometric structure).  These are 
the consistent background spacetimes for heterotic string theory.

Spacetime is kept compact and riemannian in order to keep the 2d 
model under control.  The issue of analytic continuation to real 
time is put off indefinitely.  The continuation can certainly be 
done in some extreme limits of 10d compact riemannian spacetimes, 
where at least one dimension becomes infinitely large.

\subsection*{The infrared failure of string theory}

String theory has failed as physics because of the manifold of 
consistent background spacetimes.  String theory can make no 
definite predictions, because the background spacetime depends on 
continuously variable free parameters which have to be fixed by 
hand.  Moreover, the existence of a manifold of possible background 
spacetimes implies that the low energy string modes are exactly 
massless particles.  Few, if any, real particle masses are exactly 
zero.  One of the most urgent problems in physics is to explain the 
precise small nonzero values of the particle masses in the Standard 
Model.

My purpose in pointing to the failure of string theory is 
diagnostic.  The free parameters and the massless particles are 
signs of pathology, calling for a remedy.  The pathology is in the 
infrared, its symptoms being the zero modes of the spacetime fields.  
The infrared properties of string theory are encoded in the 
background spacetimes.  Something is wrong or missing in the 
specification of the possible background spacetimes.  The need to 
select a background spacetime is the basic sign of deficiency.  The 
background spacetime should not have to be selected by hand, from 
among a manifold of consistent possibilities.  It should be produced 
mechanically.  A mechanism is missing that constructs the background 
spacetime.  One might hope that such a mechanism will remedy the 
infrared failure of string theory.

\subsection*{Every handle diverges logarithmically}

\begin{figure}
\begin{center}
\includegraphics[scale=.7]{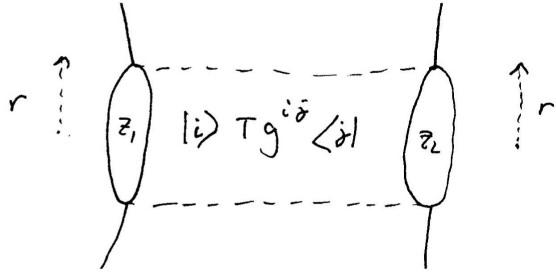}
\caption{States flowing through a handle}
\label{fig_handle}
\end{center}
\end{figure}

Consider a handle in the worldsurface, as pictured in figure 
\ref{fig_handle}.  The handle is made by cutting two disks of radius 
$r$ from the worldsurface, then gluing the two boundary circles to 
each other.  The positions $z_{1}$ and $z_{2}$ of the holes, and the 
radius $r$, are integration variables in the formula for string 
scattering amplitudes.  The effect of the handle is to make a 
non-local insertion in the worldsurface,
\begin{equation}
\label{eqn:handle}
\frac12 
\int \dif^{2}z_1\,\mu^{2} \frac1{2\pi}
\;\phi_{i}(z_1,\bar z_1)
\;
\int \dif^{2}z_2\,\mu^{2} \frac1{2\pi}
\; \phi_{j}(z_2,\bar z_2)
\int\limits_{\mu\Lambda^{-1}}
2\, \dif ( \mu r)
\: 
(\mu r)^{\ad(i)+\ad(j)-1}
\:
T g^{ij}
\end{equation}
which expresses the handle as a sum over the string states $i$, $j$ 
flowing through the two ends.  The state $i$ flowing through the end 
at $z_{1}$ appears in the worldsurface as the local field 
$\phi_{i}(z_{1},\bar z_{1})$.  The state $j$ appears as 
$\phi_{j}(z_{2},\bar z_{2})$.  Each pair of states $i,j$ is weighted 
by a gluing matrix element $T\,g^{ij}$.  The form of the integral is 
dictated by 2d scale invariance.  The 2d distance $\Lambda^{-1}$ 
acts as short distance cutoff on $r$.  The handle gluing matrix 
$T\,g^{ij}$ is the inverse of the metric
\begin{equation}
T^{-1} g_{ij} = \Z\expval{\,\phi_{i}(z_{1},\bar z_{1})
\, \phi_{j}(z_{2},\bar z_{2})\,} 
\end{equation}
which is the unnormalized two point function of the scaling fields, 
evaluated at the reference distance $\abs{z_{1}-z_{2}}=\mu^{-1}$ (on 
the complex plane compactified to form the 2-sphere).  The number 
$T^{-1}$ is the normalizing factor $\Z\expval{1}$, the partition 
function of the 2-sphere.  In a macroscopic spacetime of volume $V$, 
the number $T$ is related to the string coupling constant $\gst$ by
\begin{equation}
T^{-1} = \gst^{-2} \, V
\:.
\end{equation}
The factor $V$ comes from the zero mode in the worldsurface 
functional integral over $x^{\mu}(z,\bar z)$.  The combination 
$V\,g_{ij}$ is properly normalized to be the usual continuum inner 
product on the spacetime wave modes.

The kernel of the non-local insertion,
is the factor in equation~\ref{eqn:handle},
\begin{equation}
\int\limits_{\mu\Lambda^{-1}}
2\, \dif ( \mu r)
\: 
(\mu r)^{2\ad(i)-1}
\:
T g^{ij}
\end{equation}
where $\ad(i)+\ad(j)$ is replaced by $2\ad(i)$, because $T g^{ij} = 
0$ unless $\ad(i) = \ad(j)$ (another consequence of 2d scale 
invariance).  The cutoff dependent part of the kernel is
\begin{equation}
\label{eqn:cutoff_dependence}
T g^{ij}  \:\:
\left [
\frac{(\mu\Lambda_{1}^{-1})^{2\ad(i)}- 
(\mu\Lambda^{-1})^{2\ad(i)}} {\ad(i)}
\right ]
\end{equation}
where $\Lambda_{1}^{-1}$ is a second 2d distance, chosen 
arbitrarily, independent of $\Lambda^{-1}$, to be the upper limit in 
the integral over $r$.

The handle is logarithmically divergent in $\Lambda^{-1}$ if and 
only if there is at least one worldsurface field $\phi_{i}(z,\bar 
z)$ with $\ad(i)=0$.  The infrared failure of string theory, due to 
the existence of marginal coupling constants $\lambda^{i}$, now 
appears as a technical pathology of the worldsurface, a logarithmic 
divergence in every string channel due to the discrete zero modes 
flowing through every handle.

Again, the spectrum is discrete because spacetime is compact and 
riemannian.  The target manifold of the general nonlinear model was 
originally taken to be compact and riemannian so that the 2d model 
would be strictly well defined.  The action was then bounded below 
and the fluctuating field $x^{\mu}(z,\bar z)$ could not wander off 
uncontrollably in the target manifold.  The motivation is 
essentially the same now.  The background spacetime is taken to be 
compact and riemannian in order to control the spacetime infrared 
behavior of the string worldsurface, to be sure that nothing is 
missed in the spacetime infrared.

\subsection*{Potentially realistic string scattering}

When $\ad(i)$ is small, the kernel of the handle insertion becomes
\begin{equation}
T g^{ij}  \:\:
\left [
\frac{1- 
\me^{-2L^{2}\ad(i)}} {\ad(i)}
\right ]
\end{equation}
which is the string propagator, cutoff in the spacetime infrared at 
distance $L$.  The large distance modes are cut off in the infrared 
at $p(i)^{2} \approx L^{-2}$.  This is the usual demonstration of 
the well known principle that a worldsurface short distance cutoff 
is equivalent to an infrared spacetime cutoff.  The infrared cutoff 
acts systematically, in every string channel.

\begin{figure}
\begin{center}
\includegraphics[scale=.7]{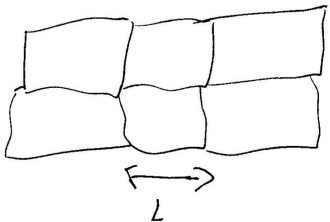}
\caption{Experimental regions of size $L$.}
\label{fig_cells}
\end{center}
\end{figure}

When the worldsurface is cut off at short 2d distance 
$\Lambda^{-1}$, the manifold of consistent background spacetimes is 
the manifold $M(L)$.  It is not necessary to satisfy $\beta=0$ 
exactly.  The non-renormalizable coupling constants $\lambda^{i}$ 
are zero in $M(L)$, so the equation $\beta=0$ is satisfied at 
spacetime distances smaller than $L$, which is all that is needed 
for consistent string scattering with infrared cutoff $L$.  The 
spacetime wave modes at distances greater than $L$ can have 
$\beta\neq 0$, so thay can act as sources and detectors for strings.  
They describe the physical environment, including the experimental 
apparatus and the observers.  Spacetime can be pictured as in 
figure~\ref{fig_cells}, tiled by cells of size $L$, which are 
experimental regions for string scattering.
The number $L$ now plays a double role.  $L$ is the IR cutoff 
distance imposed systematically on string scattering.  $L$ is also 
the UV cutoff distance in the background spacetime.  The background 
spacetime is specified by the spacetime fields at distances larger 
than $L$.  String scattering takes place within the background 
spacetime, at distances smaller than $L$.

This is a potentially realistic version of string scattering, in the 
sense that the scattering takes place at relatively small distance 
in a real physical environment.  But the large distance mechanical 
environment is still described only by classical fields.  The 
equation $\beta=0$ is still only a consistency condition.  It is now 
the tree-level condition for extending the string scattering 
amplitudes to spacetime distances larger than $L$.  Beyond the 
tree-level approximation, $\beta =0$ is not sufficient.  The cutoff 
dependence of the handles gets in the way.

\subsection*{Local handles}

\begin{figure}
\begin{center}
\includegraphics[scale=.7]{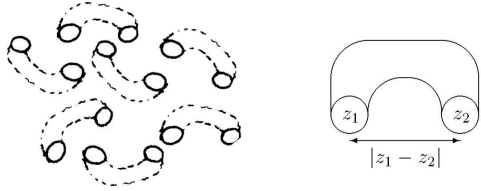}
\caption{Local handles}
\label{fig_local_handles} 
\end{center}
\end{figure}

The logarithmic divergence is interpreted as symptom of a deficiency 
in the specification of the background spacetime, a deficiency in 
the construction of the string worldsurface.  A new mechanism is to 
be added to the worldsurface, designed to cancel the logarithmic 
divergence.  But we will not try to remove the divergence in every 
string channel.  We have seen that the large distance structure of 
the background spacetime is encoded in the short distance, local 
structure of the general nonlinear model.  We limit attention, 
therefore, to the \emph{local handles}, the handles that are local 
in the worldsurface.  These are the handles that connect a local 
region of the worldsurface to itself.  Both endpoints lie within the 
same local neighborhood, as in figure \ref{fig_local_handles}.  The 
cutoff dependent effects of local handles contribute to the local 
structure of the worldsurface, and depend only on the local 
structure of the worldsurface.  They contribute to the large 
distance structure of the background spacetime, and depend only on 
the large distance structure of the background spacetime.
In limiting attention to the local handles, we keep the large 
distance physics, which is the physics of the background spacetime, 
independent from the small distance physics of string scattering.  
The principle of 2d locality is used to accomplish the separation 
systematically.  A theory of large distance physics that does not 
depend on small distance physics is attractive because it seems 
unlikely that a theory of small distance physics could be reliable.

The strategy is to add a new local mechanism at short distance on 
the worldsurface, designed precisely to cancel the divergent effects 
of the local handles.  By acting at short 2d distance, the new 
mechanism acts on the large distance structure of spacetime.  Two 
dimensional locality is preserved, so the formulas for calculating 
string scattering amplitudes will remain consistent at small 
spacetime distance even after the new mechanism is added to the 
worldsurface.

The cutoff dependence of a local handle can be extracted naturally, 
using the separation $\abs{z_{1}-z_{2}}$ between the two endpoints 
in place of the arbitrary second 2d distance $\Lambda_{1}^{-1}$ of 
equation~\ref{eqn:cutoff_dependence}.  The cutoff dependent effect 
of the local handle is the insertion
\begin{eqnarray}
\nonumber
\lefteqn{
\frac12 \int 
\dif^{2}z_1\,\mu^{2} \frac1{2\pi} \;\phi_{i}(z_1,\bar z_1) \; \int 
\dif^{2}z_2\,\mu^{2} \frac1{2\pi} \; \phi_{j}(z_2,\bar z_2)
} \\[2ex]
\label{eqn:bilocal}
& & T g^{ij}  \; 
\left [
\frac{(\mu\abs{z_{1}-z_{2}})^{2\ad(i)}
- (\mu\Lambda^{-1})^{2\ad(i)}} {\ad(i)}
\right ]
\:.
\end{eqnarray}
It is bi-local rather than non-local, in the sense that the 
insertion points $z_{1}$ and $z_{2}$ range over the same local 
neighborhood of the worldsurface.

\subsection*{Cancel the cutoff dependence of local handles}

To cancel the cutoff dependence of the local handles, insert sources
\begin{equation}
\me^{-\int \dif^{2}z \, \mu^{2} \frac1{2\pi} \,
\lambda^{i}(z,\bar z) \, \phi_i(z,\bar z)}
\end{equation}
in the worldsurface, then set them fluctuating with gaussian 
propagator
\begin{equation}
\expval{\,\lambda^i(z_1,\bar z_1) \;\; \lambda^j(z_2,\bar z_2)\,} 
=
T\,g^{ij}
\;
(\mu\Lambda^{-1})^{2\ad(i)} \;
\left [
\frac{1 - (\Lambda \abs{z_{1}-z_{2}})^{2\ad(i)}}
{\ad(i)}
\right ]
\end{equation}
which is the negative of the kernel in equation~\ref{eqn:bilocal}.  
The sources inserted in the worldsurface, when contracted with the 
$\lambda$-propagator, make bi-local insertions that cancel the 
cutoff dependence of the gas of local handles, in the independent 
handle approximation.  Interactions will have to be introduced to 
account for handle collisions.

Notice that the $\lambda$-fluctutations take place at 2d distances 
$\abs{z_{1}-z_{2}} < \Lambda^{-1}$, at the 2d distances where the 
$\lambda$-propagator is positive.  The new mechanism is acting at 2d 
distances \emph{shorter} than the characteristic distance 
$\Lambda^{-1}$.  On the other hand, the worldsurface is constructed 
at 2d distances \emph{longer} than $\Lambda^{-1}$.  The local 
handles make insertions at 2d distances $\abs{z_{1}-z_{2}} > 
\Lambda^{-1}$, at the 2d distances where the kernel in 
equation~\ref{eqn:bilocal} is positive.

Notice also that the insertions of the irrelevant scaling fields 
$\phi_{i}(z,\bar z)$ are suppressed by the appropriate factor 
$(\mu/\Lambda)^{\ad(i)}$.  There is no need for the 
non-renormalizable $\lambda^{i}$ to fluctuate.  Only the 
renormalizable $\lambda^{i}$ have to be turned into fluctuating 
local sources $\lambda^{i}(z,\bar z)$.  Thus only a finite number of 
sources $\lambda^{i}(z,\bar z)$ are needed to fluctuate at every 
short 2d distance $\Lambda^{-1}$.

The $\lambda$-propagator at 2d distance 
$\abs{z_{1}-z_{2}}\approx\Lambda^{-1}$ is
\begin{equation}
\expval{\,\lambda^i(z_1,\bar z_1) \;\; \lambda^j(z_2,\bar z_2)\,} 
\approx - T\,g^{ij}\, (\mu \Lambda^{-1})^{2\ad(i)} \;
\ln (\Lambda^{2} \abs{z_{1}-z_{2}}^{2})
\end{equation}
which is the propagator of massless scalar fields governed by a 
metric $(\Lambda \mu^{-1})^{2\ad(i)} \, T^{-1}g_{ij} $ that depends 
on the 2d distance $\Lambda^{-1}$.  After changing variables to the 
running coupling constants 
$\lambdar^{i}=(\Lambda/\mu)^{\ad(i)}\lambda^{i}$,
\begin{equation}
\lambdar^{i}(z,\bar z) = (\Lambda \mu^{-1})^{\ad(i)}
\lambda^{i}(z,\bar z)
\end{equation}
\begin{equation}
\expval{\,\lambdar^i(z_1,\bar z_1) \; \lambdar^j(z_2,\bar z_2)\,} 
\approx
- T\,g^{ij} \,
\ln (\Lambda^{2} \abs{z_{1}-z_{2}}^{2})
\:,
\end{equation}
the gaussian fluctuations of the $\lambdar^{i}(z,\bar z)$ are 
governed by the same metric $T^{-1}g_{ij}$ at every 2d distance 
$\Lambda^{-1}$.  They are produced by a functional integral over the 
$\lambdar^{i}(z,\bar z)$ with action
\begin{equation}
\int \dif^{2}z \, \frac1{2\pi}
\,
T^{-1} g_{ij}
\,
\partial \lambdar^{i}
\,
\bar \partial \lambdar^{j}
\end{equation}
which has the same form at every 2d distance $\Lambda^{-1}$.  The 
gaussian $\lambda$-fluctuations are therefore invariant under change 
of the 2d distance $\Lambda^{-1}$, when the change of scale is 
accompanied by a change of the variables $\lambda^{i}(z,\bar z)$, 
keeping the $\lambdar^{i}(z,\bar z)$ constant.

\subsection*{The $\lambda$-model}

Interactions have to be added to the gaussian action in order to 
account for collisions among the local handles.  The form of the 
interactions is constrained by 2d scale invariance and locality.  
The fields $\lambdar^{i}(z,\bar z)$ are dimensionless massless 
scalar fields.  The most general dimensionless local interactions 
are products of an arbitrary number of fields, with exactly two 
derivatives.  The action must have the form
\begin{equation}
\label{eqn:action-taylor}
S(\lambdar) = \int \dif^{2}z \, \frac1{2\pi} 
\, \left (
T^{-1} g_{ij}
\,
\partial \lambdar^{i} \, \bar \partial \lambdar^{j}
+
T^{-1} g_{ij,k} \lambdar^{k}
\,
\partial \lambdar^{i} \, \bar \partial \lambdar^{j}
+\cdots 
\right )
\end{equation}
which is that of a 2d nonlinear model.

The reference background spacetime loses its significance once the 
$\lambdar^{i}(z,\bar z)$ are set fluctuating.  Each local 2d region 
can be described just as well by fluctuations around a nearby 
background spacetime $\lambdar$.  The gaussian fluctuations around 
$\lambdar$ will cancel independent local handles in $\lambdar$.  So 
the gaussian fluctuations around $\lambdar$ have to be governed by 
the metric in $\lambdar$, $T^{-1}g_{ij}(\lambdar)$.  The 
interactions are determined completely, given the gaussian 
fluctuations around every $\lambdar$.  The action has to be
\begin{equation}
S(\lambdar) = \int \dif^{2}z \, \frac1{2\pi}
\,
T^{-1} g_{ij}(\lambdar)
\,
\partial \lambdar^{i}
\,
\bar \partial \lambdar^{j}
\:.
\end{equation}
This argument avoids direct calculation in the gas of local handles.  
The $\lambda$-model is the functional integral
\begin{equation}
\int \Dlambdar
\; \me^{- S(\lambdar)} \;
\me^{-\int \dif^{2}z \, \Lambda^{2} \frac1{2\pi} \,
\lambdar^{i}(z,\bar z) \, \phi^{\Lambda}_i(z,\bar z)}
\;.
\end{equation}
It is a 2d nonlinear model whose target manifold is $M(L)$, the 
manifold of general nonlinear models of the worldsurface.  The local 
sources $\lambdar^{i}(z,\bar z)$ are the component fields of a map 
$\lambdar(z,\bar z)$ from the worldsurface to $M(L)$.  The 
$\lambda$-model is a functional integral over the maps 
$\lambdar(z,\bar z)$.  It is inserted at short distance in the 
worldsurface to cancel the cutoff dependence of the local handles.  
The metric coupling $T^{-1}g_{ij}(\lambdar)$ is a natural 
mathematical object determined by the local properties of the 
worldsurface.  It is the inverse of the gluing matrix of a local 
handle.  Equivalently, it is the unnormalized two point function of 
the short distance fields $\phi^{\Lambda}_{i}(z,\bar z)$, evaluated 
at the short 2d distance $\Lambda^{-1}$.  The coupling strength of 
the $\lambda$-model is the string coupling constant, since $T$ is 
proportional to $\gst^{2}$.

The $\lambda$-model is scale invariant in the generalized sense.  A 
change of $\Lambda^{-1}$ is merely equivalent to a change of 
variables in the functional integral.  The action, written in terms 
of $\lambdar(z,\bar z)$, has the same form at every 2d distance 
$\Lambda^{-1}$.  When written in terms of $\lambda(z,\bar z)$, the 
action is invariant under an infinitesimal 2d scaling $ 
\Lambda^{-1}\rightarrow (1+\epsilon)\, \Lambda^{-1}$ when combined 
with the infinitesimal renormalization group transformation of the 
target manifold $M(L)$, $\lambda^{i} \rightarrow 
\lambda^{i}+\epsilon \,\beta^{i}(\lambdar)$, which is only a change 
of variables in the functional integral defining the 
$\lambda$-model.  The field $\lambdar(z,\bar z)$ is the natural 
variable in which to describe the fluctuations at the particular 2d 
distance $\Lambda^{-1}$.  The field $\lambda(z,\bar z)$ is the 
natural variable in which to describe the accumulation of 
fluctuations over a range of 2d distances $\Lambda^{-1}$, because 
the sources $\lambda^{i}(z,\bar z)$ couple to fields 
$\phi_{i}(z,\bar z)$ which are independent of $\Lambda^{-1}$.

The $\lambda$-model is a somewhat peculiar nonlinear model in that 
its target manifold $M(L)$ changes with the 2d distance 
$\Lambda^{-1}$, depending on $L^{2} = \ln (\Lambda/\mu)$.  The 
target manifold is parametrized by the coupling constants 
$\lambda^{i}$ that are renormalizable at 2d distance $\Lambda^{-1}$, 
which are the spacetime wave modes at spacetime distances larger 
than $L$.  Equivalently, the target manifold $M(L)$ is parametrized 
by the running coupling constants $\lambdar^{i}$, modulo the 
irrelevant coupling constants, which are the wave modes at distances 
smaller than $L$.  The dependence of the target manifold on 
$\Lambda^{-1}$ is unfamiliar, but poses no essential difficulty.  
The target manifold can be held fixed over any range of 2d 
distances, because extra irrelevant $\lambdar^{i}$, at distances 
somewhat smaller than $L$, can be included without noticeable 
effect.  They are decoupled from the renormalized worldsurface.  The 
dependence of the target manifold on $\Lambda^{-1}$ arranges itself 
automatically.

\subsection*{The \emph{a priori} measure}

A 2d statistical model, such as the $\lambda$-model, is made by 
accumulating fluctuations at longer and longer distances, starting 
at a very short cutoff distance $\Lambda_{0}^{-1}\approx 0$ and 
proceeding up to the characteristic distance $\Lambda^{-1}$.  Two 
pieces of data characterize the model at each 2d distance 
$\Lambda^{-1}$.  The metric coupling $T^{-1}g_{ij}(\lambdar)$ 
governs the fluctuations at distances near $\Lambda^{-1}$.  The 
\apm\ $\dif \rhor(\Lambda,\lambdar)$ is the net distribution of all 
fluctuations that have already taken place, at 2d distances up to 
$\Lambda^{-1}$.  The \apm\ is a measure on the target manifold 
$M(L)$.

The \apm\ is a familiar construct of statistical mechanics.  In the 
2d Ising model, for example, the target manifold consists of two 
points, the two possible phases of each domain.  The characteristic 
2d distance $\Lambda^{-1}$ is the lattice spacing.  The action 
governs the correlations between nearby domains.  The \apm\ is the 
statistical weight on the two possible phases at each site.  The 
\apm\ summarizes the distribution of phase domains at 2d distances 
shorter than the lattice spacing.  In models with internal 
symmetries, such as the Ising model, the \apm\ can be determined 
uniquely by the symmetry.  In general, however, the \apm\ is 
produced by the accumulation of fluctuations over the range of short 
2d distances up to $\Lambda^{-1}$.  The \apm\ evolves under the 
renormalization group of the model as fluctuations at longer and 
longer 2d distances are included.

The \apm\ is determined by its expectation values, which are 
calculated as the one point expectation values in the 
$\lambda$-model,
\begin{equation}
\int\limits_{M(L)}  f(\lambdar)
\:
\dif \rhor(\Lambda,\lambdar)
= \expval{\, f \,}
=
\expval{\, f(\lambdar(z,\bar z) ) \,}
\:,
\end{equation}
including the $\lambda$-fluctuations at 2d distances up to 
$\Lambda^{-1}$.  The renormalization group of the $\lambda$-model 
acts as a diffusion process,
\begin{equation}
-\Lambda \partialby{\Lambda} \, \dif \rhor
=
\del_{i} 
\left [
T\,g^{ij}(\lambdar) \del_{j} + \beta^{i}(\lambdar)
\right ]
\,
\dif \rhor
\end{equation}
with diffusion ``time'' being the logarithm of the 2d distance.  The 
measure diffuses in $M(L)$ because more and more fluctuations are 
included as $\Lambda^{-1}$ increases.  At the same time, $M(L)$ 
flows under the renormalization group of the general nonlinear 
model, so the diffusion is driven inward along the vector field 
$-\beta^{i}(\lambdar)$ towards the $\beta = 0$ manifold.

At tree-level, the coefficients of the diffusion process are the 
metric coupling $T^{-1}g_{ij}(\lambdar)$ and the worldsurface 
$\beta$-function $\beta^{i}(\lambdar)$.  These are constant in 
``time,'' because the $\lambda$-model is classically scale 
invariant, in the generalized sense.  A diffusion process with 
constant coefficients tends to equilibrium.  The \apm\ will diffuse 
to an equilibrium measure under the renormalization group of the 
$\lambda$-model.  The conditions that determine the \apm\ are the 
conditions of equilibrium, rather than symmetry conditions.  The 
equilibrium measure can be determined explicitly, because 
$\beta^{i}(\lambdar)$ is the gradient of a potential function 
$T^{-1} a(\lambdar)$ on $M(L)$,
\begin{equation}
\beta^{i} (\lambdar) = 
T\, g^{ij} \; 
\partial_{j} \left ( T^{-1} a(\lambdar) \right )
\:.
\end{equation}
When a diffusion process is driven by a gradient flow, the 
equilibrium measure solves the first order equation
\begin{equation}
\label{eqn:eqn_of_motion}
0=
\left [
T\,g^{ij} \del_{j} + \beta^{i}
\right ]
\,
\dif\rhor
\end{equation}
whose solution is
\begin{equation}
\label{eqn:action_principle}
\dif\rhor(\lambdar) = \me^{-T^{-1}a(\lambdar)} \; \dvol(\lambdar)
\end{equation}
where $\dvol(\lambdar)$ is the metric volume element.

The \apm\ of the $\lambda$-model is a measure on the spacetime wave 
modes $\lambda^{i}$, a functional integral over the spacetime 
fields.  It is a quantum field theory in spacetime.  Its equation of 
motion is $\beta=0$, as expressed by the equilibrium condition, 
equation~\ref{eqn:eqn_of_motion}.  Its action is the potential 
function $T^{-1}a(\lambdar)$, as shown in 
equation~\ref{eqn:action_principle}.  This QFT governs the large 
distance physics in spacetime, because the \apm\ of the 
$\lambda$-model governs the short distance structure of the 
worldsurface.  Thus the $\lambda$-model mechanically produces a QFT 
with equation of motion $\beta = 0$, which governs the large 
distance physics in spacetime.

The expectation values $ \expval{\, \lambdar^{i_{1}} 
\lambdar^{i_{2}} \cdots \,} $ in the \apm\ are the correlation 
functions of the {QFT}.  Each index $i$ represents a spacetime wave 
vector $p(i)$, along with whatever other properties label the 
spacetime wave mode $\lambda^{i}$, so the expectation values in the 
\apm\ are the momentum space correlation functions of the spacetime 
QFT. For example, at tree level,
\begin{eqnarray}
\expval{\, 
\lambdar^{i} \, \lambdar^{j}\,} &=& \expval{\, 
\lambdar^{i}(z_{1},\bar z_{1}) \, \lambdar^{j} (z_{2},\bar 
z_{2})\,}_{/z_{1}=z_{2}} \nonumber \\[2ex]
&=& T\,g^{ij} \; \left [ 
\frac{1 - (\Lambda \abs{z_{1}-z_{2}})^{2\ad(i)}} {\ad(i)} \right 
]_{/z_{1}=z_{2}} \\[1ex]
&=&T\,g^{ij}\;\frac{1}{\ad(i)} \nonumber 
\end{eqnarray}
which is the tree level spacetime propagator.  The expectation 
values in the \apm\ can be transcribed directly into spacetime 
correlation functions by summing over the spacetime wave modes, as 
in
\begin{equation}
\expval{\,h_{\mu_{1}\nu_{1}}(x_{1})\;
h_{\mu_{2}\nu_{2}}(x_{2}) \,}
 =
\expval{\, \lambdar^{i_{1}} \lambdar^{i_{2}} \,}
\;
\delta_{i_{1}}h_{\mu_{1}\nu_{1}}(x_{1})
\;
\delta_{i_{2}}h_{\mu_{2}\nu_{2}}(x_{2})
\:.
\end{equation}
All of this structure carries over from the tree-level approximation 
to the full quantum $\lambda$-model.  The fluctuations in the 
$\lambda$-model correct the metric coupling $T^{-1}g_{ij}$ and the 
$\beta$-function $\beta^{i}$, producing an effective metric coupling 
$T^{-1}\geff_{ij}$ and an effective $\beta$-function $\betaeff^{i}$.  
I argue in \cite{tentative} that the quantum $\lambda$-model is 
scale invariant, in the generalized sense, and that $\betaeff^{i}$ 
is the gradient of a potential function $T^{-1}\aeff$.  The full 
\apm\ of the quantum $\lambda$-model is generated by the diffusion 
process whose constant coefficients are $T^{-1}\geff_{ij}$ and 
$\betaeff^{i}$.  The actual equilibrium \apm\ is a QFT whose action 
is $T^{-1}\aeff$ and whose equation of motion is $\betaeff = 0$.

\subsection*{Nonperturbative 2d effects in the $\lambda$-model?}

The $\lambda$-model is merely a somewhat elaborate 2d nonlinear 
model.  It seems reasonable to presume that it can be constructed 
nonperturbatively.  Nonperturbative effects in the $\lambda$-model 
would be nonperturbative in the string coupling constant $\gst$, 
because $T\propto \gst^{2}$.  Nonperturbative 2d effects in the 
$\lambda$-model might make useful corrections to the spacetime 
equation of motion $\betaeff=0$ and to the spacetime action 
$T^{-1}\aeff$.  The most obvious source of nonperturbative 2d 
effects are the nontrivial classical solutions of the 2d equation of 
motion
\begin{equation}
0 = \partial
\left ( T^{-1}g_{ij}(\lambdar)
\, \bar \partial \, \lambdar^{j} \right )
\:.
\end{equation}
These are the harmonic surfaces in the manifold of spacetime fields.  
Solutions on the 2-sphere would be 2d instantons in the 
$\lambda$-model.  Harmonic surfaces passing through singular points 
in $M(L)$ would appear as defects in the worldsurface.  Condensation 
of such 2d defects in the $\lambda$-model might have interesting 
consequences, among which perhaps making spacetime macroscopic and 
dynamically inducing spacetime topology change.  The crucial 
immediate question is whether there are 2d nonperturbative effects 
in the $\lambda$-model can eliminate the continuous degeneracy in 
the manifold of background spacetimes, violate spacetime 
supersymmetry, and produce realistic small particle masses.

\subsection*{General covariance and gauge invariance}

The renormalization of the general nonlinear model is generally 
covariant in its target manifold, which is spacetime.  The 
reparametrizations of spacetime act on the wave modes of the 
spacetime fields, which are the coupling constants $\lambda^{i}$.  
The renormalization of the general nonlinear model is invariant, as 
long as no anomalies intervene.  The renormalization is similarly 
invariant under spacetime gauge transformations.

In the $\lambda$-model, the reparametrizations of spacetime and the 
spacetime gauge transformations become symmetries of \emph{its} 
target manifold, $M(L)$.  These symmetries of $M(L)$ are made into 
local 2d gauge symmetries in the $\lambda$-model, by introducing 
auxiliary spin-1 $\lambda$-fields coupled to the spin-1 worldsurface 
scaling fields, which are the generators of the spacetime gauge 
transformations.  The renormalization of the $\lambda$-model needs 
to preserve these symmetries of its target manifold.  In the general 
nonlinear model, the renormalization of special symmetries of the 
target manifold followed from the general covariance of the 
renormalization.  In the $\lambda$-model, the renormalization of the 
symmetries of \emph{its} target manifold should follow from the 
general covariance of \emph{its} renormalization.  General 
covariance in the target manifold of the $\lambda$-model is 
reparametrization invariance in the manifold of spacetime fields, a 
sort of \emph{meta} general covariance.  It must be shown that no 
anomalies intervene.  The same formal technology that was used in 
the renormalization of the general nonlinear model should apply to 
the $\lambda$-model.  If the renormalization of the $\lambda$-model 
is generally covariant in its target manifold, then its \apm, the 
spacetime QFT, will be generally covariant and gauge invariant in 
spacetime.

\subsection*{The effective worldsurface}

The $\lambda$-fluctuations act as fluctuating sources, making 
insertions in the worldsurface at 2d distances shorter than 
$\Lambda^{-1}$.  These insertions produce corrections to the general 
nonlinear model.  They produce an effective model of the 
worldsurface, at 2d distances longer than $\Lambda^{-1}$.  It is 
invariant under an effective renormalization group, generated by the 
effective $\beta$-function $\betaeff^{i}$.  The effective 
worldsurface, cut off at short 2d distance $\Lambda^{-1}$, is used 
to calculate effective string scattering amplitudes at spacetime 
distances smaller than $L$.  This effective model of the 
worldsurface is the effective background spacetime for string 
scattering at spacetime distances smaller than $L$.

\subsection*{The form of the theory}

For each large value of the number $L$, the $\lambda$-model produces 
two complementary effective descriptions of spacetime physics: the 
\apm, a specific QFT at spacetime distances larger than $L$, and the 
effective worldsurface, a specific effective background spacetime 
for string scattering at distances smaller than $L$.  The two 
descriptions agree where they overlap, at distances on the order of 
$L$.  There is no single description that applies at all spacetime 
distances.

The two complementary descriptions are produced by the 
$\lambda$-model working downwards in $L$.  The $\lambda$-model 
starts at some extremely short 2d distance $\Lambda_{0}^{-1}\approx 
0$ and works outward to $\Lambda^{-1}$.  The characteristic 
spacetime distance slides downwards from $L_{0}\approx\infty$ to 
$L$.  There is no dependence on the 2d cutoff distance 
$\Lambda_{0}^{-1}$ because the $\lambda$-model is renormalizable, as 
a 2d nonlinear model.  The $\lambda$-model works entirely at large 
distance in spacetime.  QFT is not derived from a microscopic, small 
distance mechanics.  Rather, it is constructed from the largest 
distances \emph{downwards}.  Such a method of producing QFT might 
turn out useful, given the need to explain the mysteriously small 
value of the cosmological constant, which is not natural in any 
microscopic {QFT}.

When the QFT is constructed downwards in distance, spacetime 
locality has to be demonstrated.  I argue that spacetime locality is 
built into the \apm\ of the $\lambda$-model by the diffusion process 
that produces it.  Locality is tested in the QFT by integrating out 
the small distance wave modes.  In the $\lambda$-model, small 
distance wave modes $\lambda^{i}$ start as non-renormalizable 
coupling constants, decoupled from the wave modes at larger 
distances and forced to zero.  As $\Lambda^{-1}$ increases, as $L$ 
decreases, the wave modes at spacetime distances near $L$ become 
renormalizable couplings.  They diffuse away from zero to their 
equilibrium distribution.  Integrating out the small distance wave 
modes $\lambda^{i}$ undoes that diffusion.  The \apm\ at larger 
values of $L$ is recovered from the \apm\ at smaller values of $L$ 
by integrating out the small distance wave modes.  This is spacetime 
locality in the QFT.

The renormalization of the general nonlinear model is valid only as 
long as $L^{2}=\ln(\Lambda/\mu)$ is large enough to be treated as a 
divergence.  Therefore the spacetime distance $L$ must stay large.  
It seems reasonable to suppose that $L^{2}>10^{24}$ or perhaps 
$L^{2}>10^{20}$ will be large enough.  If so, and if the unit of 
distance is within a few orders of magnitude of the Planck length, 
then $L$ can be taken smaller than the smallest distance 
experimentally observable in practice.  The \apm\ describes all 
spacetime physics at distances larger than $L$, so it will describe 
all observable physics, if this theory turns out to be right.

\subsection*{Dynamical background independence}

\begin{figure}
\begin{center}
\includegraphics[scale=.7]{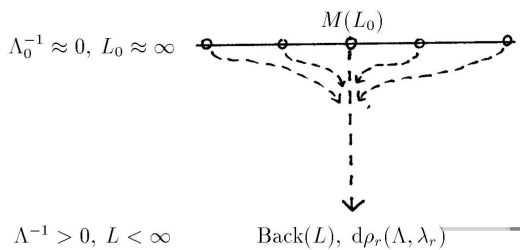}
\caption{Dynamical background independence}
\label{fig_diffuse}
\end{center}
\end{figure}

Figure~\ref{fig_diffuse} pictures schematically the production of 
the \apm\ and the effective background spacetime, which is written 
$\Back(L)$.  The process starts at $\Lambdazero^{-1}\approx 0$, 
$L_{0}\approx\infty$, in a background spacetime arbitrarily chosen 
from the manifold $M(L_{0})$.  The initial choice of background 
spacetime does not matter, because the diffusion process forgets its 
initial condition.  The $\lambda$-model is background independent, 
dynamically.

This formal argument rests on the intuition that the $\lambda$-model 
is merely a well-behaved scale invariant 2d statistical system.  It 
is presumed to behave as if its target manifold $M(L)$ is globally 
finite dimensional, compact and connected, in the limit 
$L\rightarrow \infty$.  The diffusion of the \apm\ will then go to a 
unique equilibrium.  This is a very strong presumption.  There will 
have to be nonperturbative 2d effects in the lambda model that 
regularize the singularities in $M(L)$, especially the places where 
spacetime becomes macroscopically large and the dimension of $M(L)$ 
grows without bound.  There will have to be 2d effects in the lambda 
model that change the spacetime topology.  It can be imagined that 
fluctuating 2d defects in the $\lambda$-model, associated with the 
singularities in $M(L)$, will regularize the singularities.  But 
such effects are now only imagined.

\subsection*{What the cancelling means}

The $\lambda$-model was introduced to cancel the dependence on 
$\Lambda^{-1}$ of the local handles.  But the local handles act at 
2d distances longer than $\Lambda^{-1}$, while the 
$\lambda$-fluctuations act at 2d distances shorter than 
$\Lambda^{-1}$.  It cannot be that the $\lambda$-fluctuations 
actually cancel the local handles, leaving no net effect.  Rather, 
the incremental effects of the $\lambda$-fluctuations are opposite 
to those of the local handles, as $\Lambda^{-1}$ changes.  Consider 
a small change, $ \Lambda^{-1} \rightarrow 
(1+\epsilon)\Lambda^{-1}$.  The $\lambda$-fluctuations at 2d 
distances between $\Lambda^{-1}$ and $(1+\epsilon)\Lambda^{-1}$ act 
on $\Back(L)$ to produce the effective background spacetime 
$\Back(L-\delta L)$, where $\delta L =\epsilon/{2 L}$.  
Schematically,
\begin{equation}
\Back(L) + (\lambda{-}\mathrm{fluctuations})
= \Back(L-\delta L)
\:.
\end{equation}
The cancelling between the $\lambda$-fluctuations and the local 
handles at 2d distances between $\Lambda^{-1}$ and 
$(1+\epsilon)\Lambda^{-1}$ is, schematically,
\begin{equation}
(\lambda{-}\mathrm{fluctuations}) + (\mathrm{local\ handles})  =   0
\:.
\end{equation}
Therefore,
\begin{equation}
\Back(L) = \Back(L-\delta L) + (\mathrm{local\ handles})
\:.
\end{equation}
The local handles act on $\Back(L-\delta L)$ to produce $\Back(L)$.  
The local handles \emph{undo} the work of the $\lambda$-model.

In a macroscopic spacetime, with $n$ macroscopic dimensions of 
volume $V$, the discrete sums over wave modes become continuous 
momentum integrals
\begin{eqnarray}
\sum_{i,j} T g^{ij} &=& \sum_{p(i)} 
\sum_{p(j)} \;\gst^{2} \, V \;\delta_{p(i),p(j)} \\[2ex]
&=& 
\int \dif ^{n}p(i) \int \dif^{n} p(j) \; \;
\gst^{2} \; \delta^{n}(p(i)-p(j))
\end{eqnarray}
writing here only the wave vector $p(i)$ for $i= (p(i), \cdots)$,
leaving out all the other labelling of the wave modes.
The 
kernel of the handle insertion (the string propagator)
becomes
\begin{equation}
\delta^{n}(p(i)-p(j))\;\;
\gst^{2}
\; \left [
\frac{1- 
(\mu\Lambda^{-1})^{2p(i)^{2}}}{p(i)^{2}}
\right ]
\end{equation}
Take $\Lambda\partial/\partial\Lambda$ of this kernel
to find that its 2d scale dependence
\begin{equation}
\delta^{n}(p(i)-p(j))\;\;
\gst^{2}
\; \left [
2\; \me^{-2L^{2}p(i)^{2}}
\right ]
\end{equation}
is entirely at spacetime distances larger than $L$.

As the infrared cutoff is relaxed, as $L$ is increased, more and 
more large distance string modes pass through the local handles, 
modifying the background spacetime at distances larger than $L$.  As 
$L$ increases, the background spacetime evolves, running back 
through the effective background spacetimes that were produced by 
the $\lambda$-model.  The diffusion process runs backwards.  But a 
diffusion process can run backwards only on states that have first 
been produced by the forward process.  Thus string scattering at 
finite distances $L$ in spacetime is consistent only in the 
effective background spacetime produced by the $\lambda$-model.

\subsection*{Physics at finite spacetime distance}

The $\lambda$-model is, in principle, a realistic theory.  By 
working at nonzero $\Lambda^{-1}$, it produces a mechanical 
description of spacetime at large finite distances $L$.  It produces 
a QFT, an actual functional integral over spacetime fields, which 
describes the large distance physics.  It produces an effective 
background spacetime for string scattering at every large distance 
$L$, evolving consistently with increasing $L$.  String scattering 
takes place at finite distances in the mechanical large distance 
environment produced by the $\lambda$-model.  States of the QFT can 
describe mechanical string scattering experiments which probe small 
distance physics, at least hypothetically.

Pure S-matrix string theory, on the other hand, describes strings 
scattering at infinite distance in an infinitely large spacetime.  
It is an idealized theory of perfectly asymptotic states, exactly 
on-shell.  It assumes an ideal experimenter at infinity in 
spacetime.  It has no room for mechanics.  It has no room for a 
mechanical description of the physical environment in which 
scattering takes place, no room for a mechanical description of the 
experimental apparatus or the observers.  The worldsurface is in the 
idealized continuum limit, $\Lambda^{-1}=0$, at an exact solution of 
$\beta =0$.  The logarithmic dependence on $\Lambda^{-1}$ in the 
local handles is pushed away to $L=\infty$, outside the idealized 
scattering region, invisible in the S-matrix.  There are no discrete 
zero modes to produce logarithmic divergences.  (Logarithmic 
divergences can appear only in tadpole diagrams, where a non-local 
handle is attached at one of its ends to a vacuum diagram, which can 
make a contribution at the end of the handle proportional to 
$\delta^{n}(p)$.  These are \emph{not} contributions to the 
background spacetime, since they come from non-local handles.  They 
see the small distance string modes circulating in the vacuum 
diagram.)

The dependence of local handles on $\Lambda^{-1}$ is a signficant 
difficulty only in a potentially realistic theory of string 
scattering, which strives to describe scattering at finite spacetime 
distances $L$ within a mechanical environment.  Taking the 
background spacetime to be compact and riemannian distills that 
difficulty down to the logarithmic divergence in local handles, 
coming from spacetime zero modes.  Treating this logarithmic 
divergence is a way to treat the real difficulty, which is the 
dependence of the local handles on $\Lambda^{-1}$, at finite $L$.  
That difficulty cannot be avoided if string theory is to have a 
relation to the physics of the real world, because the real world 
\emph{is} mechanical at large distances.  There has to be a 
mechanical background spacetime at large finite distances $L$, which 
evolves consistently with $L$.  String theory needs the 
$\lambda$-model to prepare the effective background spacetime within 
which string scattering at finite spacetime distance is consistent.

The conventional background spacetimes for pure S-matrix string 
theory are suspect as guides to the properties of the effective 
background spacetime.  The conventional background spacetimes, the 
exact solutions of $\beta =0$, are idealized settings for pure 
string S-matrices.  Properties such as exact spacetime supersymmetry 
might turn out to be artifacts of prematurely setting 
$\Lambda^{-1}=0$, $L=\infty$.  A potentially physical S-matrix 
should derive from a theory which can describe finite mechanical 
experiments.  Only in the limiting case of infinitely large 
experimental apparatus does an S-matrix emerge.  This is how the 
S-matrix is derived from ordinary quantum field theories, such as 
the standard model.

The $\lambda$-model starts at $L_{0}\approx\infty$.  It starts at a 
background spacetime in $M(L_{0})$, which is a solution of $\beta=0$ 
at distances smaller than $L_{0}$, so is essentially a conventional 
background spacetime.  The renormalization group of the 
$\lambda$-model then runs, eliminating dependence on 
$\Lambdazero^{-1}$, forgetting the conventional background 
spacetime.  The \apm\ reaches equilibrium by the time any finite 
value of $L$ is reached.  Only after this process has taken place, 
only after 2d universality has set in, does it make sense to explore 
the limit $\Lambda^{-1}\rightarrow 0$, $L\rightarrow \infty$.  The 
nature of that limit would become the most basic question, if this 
theory should turn out to be right.

\subsection*{What needs to be done}

The $\lambda$-model is a certain 2d nonlinear model.  It is 
formulated, and its structure described, in abstract, general terms.  
The task now is to figure out what it actually does.  This is a 
familiar kind of challenge in theoretical physics, to figure out 
what is to be calculated in a model, and how to do the calculations.

The general arguments for how the $\lambda$-model works need to be 
checked in detail.  A first check can be made by calculating the one 
loop corrections in the lambda model, the corrections to the 
effective metric, the effective $\beta$-function, the \apm, and the 
effective worldsurface.  It should be possible to check explicitly 
the generalized scale invariance of the $\lambda$-model, and the 
correspondence between the \apm\ and the effective string scattering 
amplitudes.  One particular technical point to check is the equation 
$R_{ij}= g_{ij}/4$ on the graded manifold of supersymmetric 
background spacetimes, which follows from these general principles.

The most urgent task is to figure out the leading nonperturbative 
effects in the $\lambda$-model.  In particular, do nonperturbative 
2d effects remove spacetime supersymmetry?  Do they generate small 
nonzero particle masses?  Do they make spacetime macroscopic?  If 
these questions have positive answers, it will become worthwhile to 
look for a definite, detailed explanation of the Standard Model and 
the rest of our current knowledge of the real physical world, to 
find out if the $\lambda$-model is physically useful.  For now, it 
is only a speculative new way to do theoretical physics.

\subsection*{Acknowledgments}

I thank the organizers of the Carg\`ese 2002 ASI for inviting me to 
talk.

\end{document}